\documentclass[11pt]{article}
\usepackage{graphicx}
\usepackage{float}
\usepackage{verbatim} 
\usepackage{amsmath}
\usepackage{booktabs} 
\usepackage{tabularx}
\usepackage[numbers]{natbib} 
\usepackage{hyperref} 

\begin{document}

\title{Concentration in Governance Control Across Decentralised Finance Protocols}

\author{
Thomas Eisermann$^a$\thanks{Email: teisermann@fc.ul.pt} \and
Carlo Campajola$^{b,c,d}$\thanks{Corresponding author. Email: c.campajola@ucl.ac.uk} \and
Claudio J. Tessone$^{d,e}$\thanks{Email: tessone@ifi.uzh.ch} \and
Andreia Sofia Teixeira$^{f,a}$\thanks{Email: sofia.teixeira@nulondon.ac.uk}
}

\date{}

\maketitle

\begin{center}
\small
$^a$ LASIGE, Faculdade de Ciências, Universidade de Lisboa, Portugal \\
$^b$ Institute of Finance and Technology, University College London, United Kingdom\\
$^c$ DLT Science Foundation, London, United Kingdom \\
$^d$ UZH Blockchain Center, University of Zurich, Switzerland\\
$^e$ Blockchain \& Distributed Ledger Technologies Group, University of Zurich, Switzerland \\
$^f$ Network Science Institute, Northeastern University London, United Kingdom

\end{center}

\begin{abstract}
Blockchain-based systems are frequently governed through tokens that grant their holders voting rights over core protocol functions and funds. The centralisation occurring in Decentralised Finance (DeFi) protocols' token-based voting systems is typically analysed by examining token holdings' distribution across addresses. In this paper, we expand this perspective by exploring shared token holdings of addresses across multiple DeFi protocols. We construct a Statistically Validated Network (SVN) based on shared governance token holdings among addresses. Using the links within the SVN, we identify influential addresses that shape these connections and we conduct a post-hoc analysis to examine their characteristics and behaviour. Our findings reveal persistent influential links over time, predominantly involving addresses associated with institutional investors who maintain significant token supplies across the sampled protocols. Finally, we observe that token holding patterns and concentrations tend to shift in response to speculative market cycles.  
\end{abstract}


\section{Introduction \label{introduction}}
Decentralised Finance (DeFi) encompasses a family of services and protocols that replicate traditional financial functions -- such as collateralised lending and asset trading -- on permissionless, distributed ledger technology (DLT) networks. These functions are executed through smart contracts: automated software that any DLT user can operate, ensuring transactions remain transparent, irreversible, and free from reliance on third parties \citep{werner2022sok}. This architecture grants DeFi its non-custodial nature, eliminating the need for intermediaries to act as custodians of funds and assets, thereby reducing counterparty risk \citep{schar2021decentralized}.

Although the execution of smart contracts is inherently permissionless and decentralised, the configuration and governance of these functions are determined by entities with governance rights over the protocols. Governance in DeFi typically begins under the centralised control of a ``benevolent dictator" or a small council, transitioning over time to a more decentralised framework. This shift is facilitated through governance tokens, which grant holders the ability to propose or vote on protocol changes. These tokens confer membership in a Decentralised Autonomous Organisation (DAO), which manages protocol operations. Upgrades are presented as executable code, with predefined thresholds of token ownership required to propose changes and a quorum needed for their approval. While the rights associated with governance tokens vary across projects, they generally provide significant decision-making authority over protocol governance. \citep{werner2022sok}

Since a governance token represents decision-making power, the token holder distribution is a relevant factor - among others - to determine the extent of decentralisation of a protocol \citep{Axelsen2022}. While previous studies have analysed token distribution within individual protocols \citep{nadler2020decentralized, Barbereau2022, fritsch2022analyzing, Jensen2023}, few have explored the extent to which actors wield influence across multiple protocols simultaneously.

Users interact with DLT systems through unique identifiers known as addresses. These addresses are derived from public-key cryptography, where the owner holds a private key, and the public key is shared across the network. Addresses allow users to validate transactions, interact with smart contracts, such as those in DeFi applications, and enable transparent tracking of token ownership within the system.

In this paper, we inspect the dynamics of cross-protocol governance by analysing the ownership structures of governance tokens within DeFi. Governance token holders may be individuals, companies, associations, or even smart contracts. Therefore, we investigate the presence and impact of influential user groups who may hold disproportionate influence across different DeFi protocols. In our analysis we investigate the existence of such groups, who those users are, how they behave and the degree of control they may exert within the DeFi ecosystem.

\section{Background}\label{literature_review}
Our work builds on existing research addressing the emergence of centralisation in DLTs and the structure of DeFi governance, incorporating methodological approaches from complex financial network analysis.

\subsection{Emerging centralisation in DLT-based systems}

Decentralisation is often seen as a key value in Web 3.0 projects aiming to reshape power dynamics \citep{bodo2021decentralisation}. However, its definition varies depending on whether decentralisation is assessed through technological, economic, geographical, or social and governance metrics. A substantial body of research has explored the centralisation of wealth in cryptocurrency markets. For instance, \cite{makarov2021blockchain} identifies growing wealth inequality on the Bitcoin blockchain, driven primarily by the specialization of entities and the emergence of a financial intermediation industry. Similar trends have been observed in other cryptocurrency systems \citep{campajola2022evolution,campajola2022microvelocity}.

Centralisation has also been measured in technological terms by looking at the peer-to-peer infrastructure \citep{grundmann-icbc22,gao2024heterogeneity}, where emerging hubs constitute potential vulnerabilities, or at the consensus protocols of layer-1 blockchains \citep{li2023reward,brown2023measuring}, where very few mining and staking pools have accrued the majority of the validation authority.

Geographic centralisation has been another focus of study. In \cite{sun2022spatial} the authors investigate the geographic distribution of Bitcoin miners, finding that while operations span as many as 139 countries, the majority of computing power is concentrated in just a few locations.

In summary, DLTs exemplify the natural tendency of open and loosely regulated markets to concentrate power and evolve toward oligopolistic structures \citep{Arthur1994IncreasingEconomy}. This is further complicated by the limited accountability of central entities, often shielded by the anonymity granted by the design of these systems \citep{walch2019deconstructing}.

\subsection{DeFi Governance Research}\label{s-lit_review:govt_research}
The literature on DeFi governance has sought to evaluate decentralisation across its various dimensions, ranging from technical architecture to social layers and communities associated with these systems \citep{sai2021taxonomy, Axelsen2022, ruetschi2024decentralized, gogol2024sok}. Since governance tokens typically confer voting rights to their holders, several studies have focused on the distribution of these tokens and the corresponding voting power within DeFi communities to determine who ultimately controls the protocols \citep{nadler2020decentralized, fritsch2022analyzing, feichtinger2023hidden, Jensen2023}.

The study by \cite{nadler2020decentralized} examines various types of tokens, including governance tokens within DeFi, and reveals a pronounced centralisation in their ownership structure, with often 5 to 10 addresses holding more than 50\% of the supply. They also find significant variation in multi-token holding patterns depending on the specific token. Similarly, in \cite{fritsch2022analyzing} the authors report high centralisation in governance across their sample. Their findings suggest that a few addresses wield substantial control, analogous to major shareholders in traditional corporate governance. Despite this centralisation, they observe that influential entities frequently align their votes with the broader community.

In \cite{Barbereau2022}, the authors find that despite the theoretical decentralisation in DeFi protocol, voting rights tokens tend to lead to highly centralised control. Their analysis across multiple DeFi projects highlights that even token distribution strategies designed to enable ``fair launches" have not prevented centralisation over time. \cite{feichtinger2023hidden} shows that despite their intent for inclusive decision-making, many DAOs exhibit centralised control and inefficiencies, with high costs and low community participation. \cite{Jensen2023} finds that in practice, token-based voting power can undermine the democratic ethos by enabling major holders to push through unpopular decisions. The study suggests a misalignment between declared values and actual governance, proposing a framework to evaluate emerging voting systems and their potential to address these discrepancies.\cite{kitzler2024governance} find that DAO contributors, including project owners and developers, often hold significant governance influence, with majority control in 7.54\% of DAOs and last-minute token acquisitions affecting 14.81\% of proposals. However, they observe limited evidence of contributors exerting influence across competing DAOs, despite the potential incentives for governance token holders to oppose proposals that benefit rival protocols.

This growing research highlights the unequal token distribution of governance tokens, which results in a concentration of decision-making power and, therefore, control over the project. However, to the best of our knowledge, apart from the indirect findings of \cite{nadler2020decentralized} and \cite{kitzler2024governance}, no study has explicitly investigated the phenomenon of \textit{cross-protocol control} — the influence exerted by one or more entities holding governance tokens across multiple protocols, thereby impacting several communities simultaneously.

\subsection{Financial Complex Networks}\label{s-lit_review:networks}
Network analysis has been widely applied to model financial systems, as a powerful method to make interdependencies and contagion pathways apparent across assets and markets \citep{fonseca2025analysis, siudak2022network}. Among these methods, Statistically Validated Networks (SVNs) have proven particularly effective for analysing complex financial networks, by identifying statistically significant links against a background of randomly occurring connections. This approach, introduced by \cite{tumminello2011statistically}, is useful for isolating key dependencies within bi-partite systems and has been applied in various financial contexts.

In this context, \cite{bardoscia2021physics} offers a comprehensive summary of the theory behind financial networks and their main applications, illustrating how their structures influence systemic risk and the transmission of financial shocks. Financial systems are highly interconnected, and understanding these connections is crucial for identifying points of vulnerability, especially during periods of market instability.

With particular interest, \cite{gualdi2016statistically} used SVNs to analyse portfolio overlaps and assess systemic risk. By examining binary holding matrices and accounting for individual stock position sizes, they identified significant dependencies and found that the proportion of validated links increased steadily leading up to the 2007–2008 financial crisis, showing that market participants exhibited herding behaviour that amplified the crash. 



Inspired by the methodologies of \cite{tumminello2011statistically} and \cite{gualdi2016statistically}, we adapt them to examine dependencies across DeFi protocols, focusing on shared overlaps in addresses holding governance tokens.

\section{Methodology \label{section:methodology}}
\label{methodology}

\subsection{Token Selection}
We adopt the selection criteria outlined by \cite{nadler2020decentralized} to identify the most relevant governance tokens issued by DeFi protocols. These criteria are summarized as follows:

\begin{enumerate}
    \item \textit{Governance Qualification}: the token must qualify as a governance token, granting its holders the ability to influence decisions shaping the ecosystem's rules \citep{freni2022tokenomics}. Tokens classified purely as stablecoins, utility tokens, token wrappers, or token baskets are excluded;
    \item \textit{ERC-20 Compliance}: the token must adhere to the ERC-20 standard \citep{ERC20_2015};
    \item \textit{Market and Protocol Significance}: at least one of the following conditions is satisfied at the time of data collection (December 8, 2022): a) The token has a significant circulating supply with a market capitalisation (MCAP) of over 200 million USD, b) The protocol's contracts have a total value locked (TVL), excluding vested tokens, of over 300 million USD, with at least 50\% of the TVL on Ethereum Mainnet. 
\end{enumerate}

To compile the initial token list, we consulted data sources such as CoinGecko\footnote{\url{https://www.coingecko.com/}} and DeFiLlama\footnote{\url{https://defillama.com/}}. The list was further refined manually. A final version, including justifications for any exclusions, is available in the public GitHub repository linked at the end of this paper.

\begin{table}[t]
\centering
\begin{tabular}{|p{3cm}|p{8.6cm}|}
\hline
Protocol & Contract Address \\
\hline
Uniswap & 0x1f9840a85d5af5bf1d1762f925bdaddc4201f984 \\
Aave & 0x7fc66500c84a76ad7e9c93437bfc5ac33e2ddae9 \\
Lido & 0x5a98fcbea516cf06857215779fd812ca3bef1b32 \\
Maker & 0x9f8f72aa9304c8b593d555f12ef6589cc3a579a2 \\
Curve & 0xd533a949740bb3306d119cc777fa900ba034cd52 \\
1Inch & 0x111111111117dc0aa78b770fa6a738034120c302 \\
Bitdao & 0x1a4b46696b2bb4794eb3d4c26f1c55f9170fa4c5 \\
Convex & 0x4e3fbd56cd56c3e72c1403e103b45db9da5b9d2b \\
Compound & 0xc00e94cb662c3520282e6f5717214004a7f26888 \\
dYdX & 0x92d6c1e31e14520e676a687f0a93788b716beff5 \\
Balancer & 0xba100000625a3754423978a60c9317c58a424e3d \\
Sushi & 0x6b3595068778dd592e39a122f4f5a5cf09c90fe2 \\
Yearn Finance & 0x0bc529c00c6401aef6d220be8c6ea1667f6ad93e \\
Instadapp & 0x6f40d4a6237c257fff2db00fa0510deeecd303eb \\
Aura Finance & 0xc0c293ce456ff0ed870add98a0828dd4d2903dbf \\
\hline
\end{tabular}
\caption{List of selected governance tokens from DeFi protocols used in this paper, including token names and contract addresses.}
\label{tab:token_selection}
\end{table}

The final Token Selection for this study is displayed in Table \ref{tab:token_selection}.
It is important to note that Governance tokens have different rights and obligations associated with them. In Appendix \ref{appendix:overview_governance_tokens} we provide a short description of their function, governance process and governance scope.

\subsection{Time Interval Selection}
We retrieved -- from an Ethereum Erigon Node using ethereum-etl ~\citep{ethereumetl} -- monthly tables for the addresses holding the short-listed tokens in Table \ref{tab:token_selection} between 2021-01-15 and 2022-06-15 by netting the historical token transfer events. Snapshot dates and block heights are displayed in Table ~\ref{tab:sample_intervals} in Appendix ~\ref{appendix:snapshots}. The chosen blocks are the ones added to the longest blockchain closest to 12 AM UTC on the day of each snapshot.

\subsection{Address Labels}
Token balance tables are enriched with address labels, connecting pseudonymous addresses with real-world entities, for all addresses holding more than 0.01\% of the token supply of any given token at any given point in time of our selected snapshots. Labels were collected from EtherScan \footnote{\url{https://etherscan.io/labelcloud}}, Nansen \footnote{\url{ https://www.nansen.ai/}} and Arkham \footnote{\url{ https://www.arkhamintelligence.com/}} as well as the respective protocols documentation. The data sources utilised for labelling are partly closed-source and partly open-source. To ensure the reliability of the assigned labels we triangulated between them by cross-verifying labels across the different data sources and associating an address with their most frequent label. It is worth noting that there were rare conflicts during the triangulation. We removed known burner addresses and the token supply held by these addresses from our analysis.

\subsection{Statistically Validated Network Projections}
Based on the approach by \cite{tumminello2011statistically}, we constructed a bipartite network $S = \{ N_t, N_a \}$ by grouping addresses in a node set $N_a$ and the token nodes into the other set $N_t$. A link is established between a token node $t_i \in N_t $ and an address node $a \in N_a$ if the latter holds the token associated with the former. 

As in \cite{tumminello2011statistically}, we assume a hypothesis of random connectivity between addresses and tokens, accounting for the degree heterogeneity in each set. Specifically, the probability that two tokens $t_i$ and $t_j$ share $N_{i,j}^a$ addresses by chance is given by the hypergeometric distribution: 

\begin{equation}
    P(X = N_{i,j}^a) = \frac{{ \binom{N_i^a}{N_{i,j}^a} \binom{N_a - N_i^a}{N_j^a - N_{i,j}^a} }}{\binom{N_a}{N_j^a}}
\end{equation}

where $N_i^a$ and $N_j^a$ are the number of addresses holding $t_i$ and $t_j$, and $N_a$ is the total number of addresses.

We then proceed to test the presence of the link between tokens $t_i$ and $t_j$ against the hypothesis of random connectivity. This is done by computing the p-value $p(N_{i,j}^a)$, which measures the probability of observing $N_{i,j}^a$ or more shared addresses under the null hypothesis:

\begin{equation}
    p(N_{i,j}^a) = 1 - \sum_{x=0}^{N_{i,j}^a - 1} P(X = x)
\end{equation}

To control for multiple hypothesis testing, we apply a Bonferroni correction, adjusting the threshold level $\alpha$ to $\alpha / T$, where $T$ is the number of token pairs tested. We use a threshold level of $\alpha = 0.01$ to set a hard cap on the occurrence of type I errors.

The outcome is a SVN $G_V = (N_t, E_V)$, where $N_t$ is again the set of tokens and $E_V$ is the set of statistically validated edges between tokens. Each edge $(t_i, t_j) \in E_V$ indicates a statistically significant relationship between tokens $t_i$ and $t_j$, based on the number of shared addresses that hold both tokens, as determined by the hypergeometric test and the corrected significance threshold.

\subsection{Link Analysis}

The primary objective of our research is to understand the prevalence of control exerted by addresses across DeFi protocols. As highlighted by \cite{gualdi2016statistically}, binary token holding matrices do not account for position size, which is crucial for assessing the control exerted by groups of addresses across protocols or, in their case, the concentration in financial portfolios. Validating the original weighted matrix is challenging due to the lack of an analytical null model.

To overcome this limitation, we identify all links in the SVN projections based on binary token-holding matrices. These identified links between tokens $t_i$ and $t_j$ are used to filter and identify the set of relevant addresses responsible for the formation of a given link within a token projection. We define this set of relevant addresses as the \textit{link-defining addresses}, denoted by $A_{i,j}$:

\begin{equation}
    A_{i,j} = \{ a \in N_a \mid a \text{ holds both } t_i \text{ and } t_j \}
\label{eq:link_defining_address}
\end{equation}

These link-defining addresses represent the addresses that simultaneously hold both tokens $t_i$ and $t_j$, thereby contributing to the formation of the statistically validated link between these tokens. By analysing the properties of the set $A_{i,j}$, we can further investigate the characteristics of the addresses that bridge multiple protocols through token holdings.

\subsection{Key Metrics}
We analyse the population of link-defining addresses \( A_{i,j} \) that hold at least 0.0005\% of the total token supply, i.e., \( \frac{t_j}{S_t} > 0.000005 \), for at least one of the tokens in our sample over the observed time frame for the following metrics:

\textit{Internal Influence:} The internal influence of the addresses in $A_{i,j}$ is calculated as the average fraction of the total supply for tokens $t_i$ and $t_j$ held by the addresses in $A_{i,j}$. This is expressed as:
\begin{equation}
    \textit{Internal Influence} = \frac{\sum_{a \in A_{i,j}} \left( \frac{q_a(t_i)}{s(t_i)} + \frac{q_a(t_j)}{s(t_j)} \right)}{2}
\label{eq:internal_influence}
\end{equation}

where $q_a(t_i)$ and $q_a(t_j)$ are the amounts of tokens $t_i$ and $t_j$ held by address $a$, and $s(t_i)$ and $s(t_j)$ are the total supplies of tokens $t_i$ and $t_j$, respectively.

\textit{Directional Influence:} The directional influence for a given token, say $t_i$, is the normalised percentage of the token supply held by the addresses in $A_{i,j}$:

\begin{equation}
    \textit{Directional Influence}_{t_i} = \frac{\sum_{a \in A_{i,j}} q_a(t_i)}{s(t_i)}
\label{eq:directional_influence}
\end{equation}

and similarly for a token $t_j$.

\textit{Relative Influence by Label:} This metric captures the share of internal influence held by link-defining addresses $A_{i,j}$ categorised by labels $L(a)$ in a given link $(t_i, t_j)$ of $E_V$.  For a given label $L$, the relative influence is defined as:

\begin{equation}
    \text{Relative Influence}_{L} = \frac{\sum_{a \in A_{i,j}, L(a) = L} \left( \frac{h_a(t_i)}{S(t_i)} + \frac{h_a(t_j)}{S(t_j)} \right) }{\sum_{a \in A_{i,j}} \left( \frac{h_a(t_i)}{S(t_i)} + \frac{h_a(t_j)}{S(t_j)} \right) }
\label{eq:relative_internal_influence}
\end{equation}

We also consider link size (i.e. the number of addresses holding both tokens at the end of a link), median wealth and the Gini coefficient \citep{gini1921} of influence as additional metrics.








\subsection{Permuted Control}
We evaluate the results for each metric, where indicated, using a permutation test to assess differences against a control group. The control group $A_{\text{control}}$ consists of a set of randomly selected addresses, matched in size to the link-defining address set $A_{i,j}$. 

In each permutation, we randomly shuffle the addresses from the combined set $A_{i,j} \cup A_{\text{control}}$, where $A_{\text{control}}$ is the control group. Let $\hat{m}_{A_{i,j}}^{(k)}$ and $\hat{m}_{A_{\text{control}}}^{(k)}$ represent the recalculated metric of interest for the shuffled groups in the $k$-th permutation. This process is repeated for $1000$ iterations to generate a distribution of the metric under the null hypothesis of no effect. 

The significance of the observed difference $\Delta m = m_{A_{i,j}} - m_{A_{\text{control}}}$ is evaluated by calculating its percentile rank in the distribution of $\hat{\Delta m}^{(k)} = \hat{m}_{A_{i,j}}^{(k)} - \hat{m}_{A_{\text{control}}}^{(k)}$ across the $1000$ iterations.
\section{Results}\label{results} 

\subsection{Token projection}
\begin{figure}[H]  
    \centering
    \includegraphics[width=0.8\textwidth]{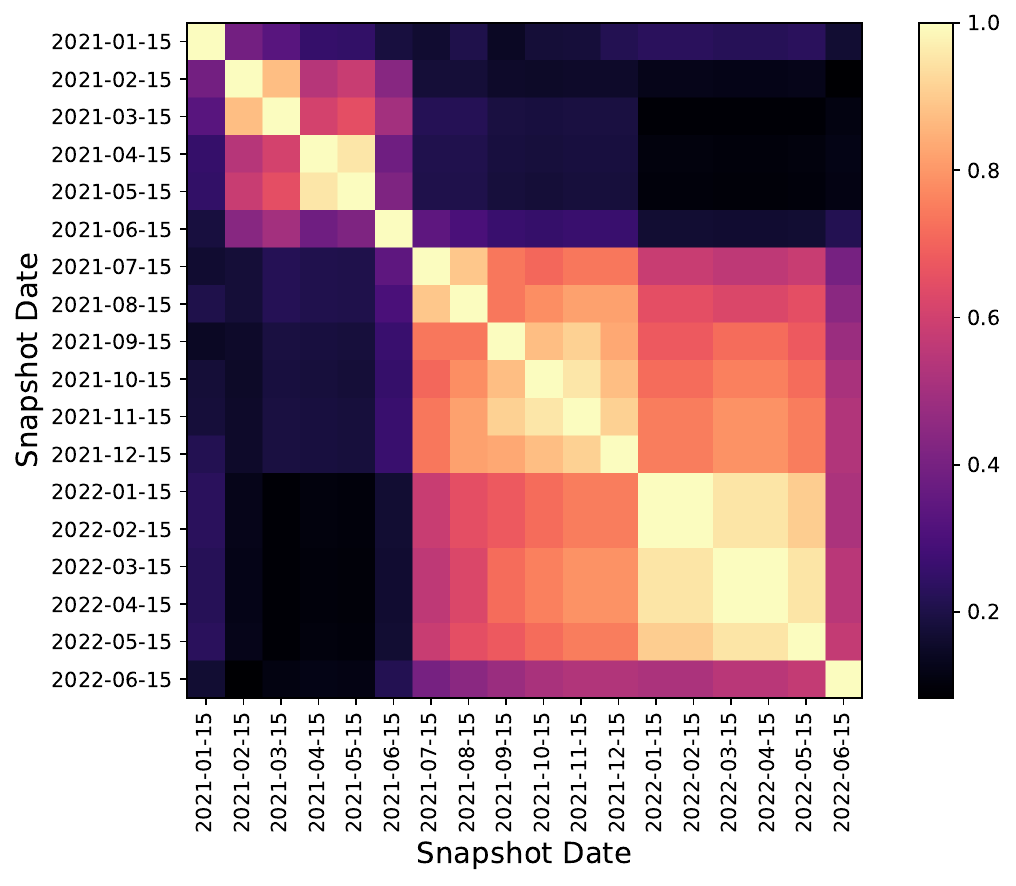}
    \caption{Jaccard similarity of the SVNs over time.} 
    \label{fig:jaccard_similarity}
\end{figure}

To understand the persistence of network links over time, we constructed a Jaccard Similarity Matrix, shown in Figure \ref{fig:jaccard_similarity}. In this figure the diagonal elements represent perfect similarity (1.0), while off-diagonal elements indicate the degree of similarity between SVN network snapshots at different points in time. 

The chart shows a relatively similar network structure of SVN snapshots before June 2021 and after June 2021. The break in similarity can be explained by the introduction of new governance tokens into our sample dataset, leading to the formation of additional validated links. We refer the reader to Figure \ref{fig:address_holding_absolute_over_time} in the Appendix, showing the absolute number of addresses holding a governance token over time and illustrating their introduction and growth in our sample. Despite the break in similarity, Figure \ref{fig:jaccard_similarity} highlight that overall links constituting the SVNs remain persistent and stable over time, resulting in high similarity of the network typology. This persistence validates our method for identifying and analysing these links and suggests that subsets of addresses contributing to the links are likely the same over time.

\subsection{Link Analysis}
Building on the persistence of links observed in the SVNs, we now examine the nature of these validated links to understand the characteristics, behaviour and potential influence exercised by set of addresses constituting the links.

\begin{figure}[H]
    \centering
    \includegraphics[width=0.8\textwidth]{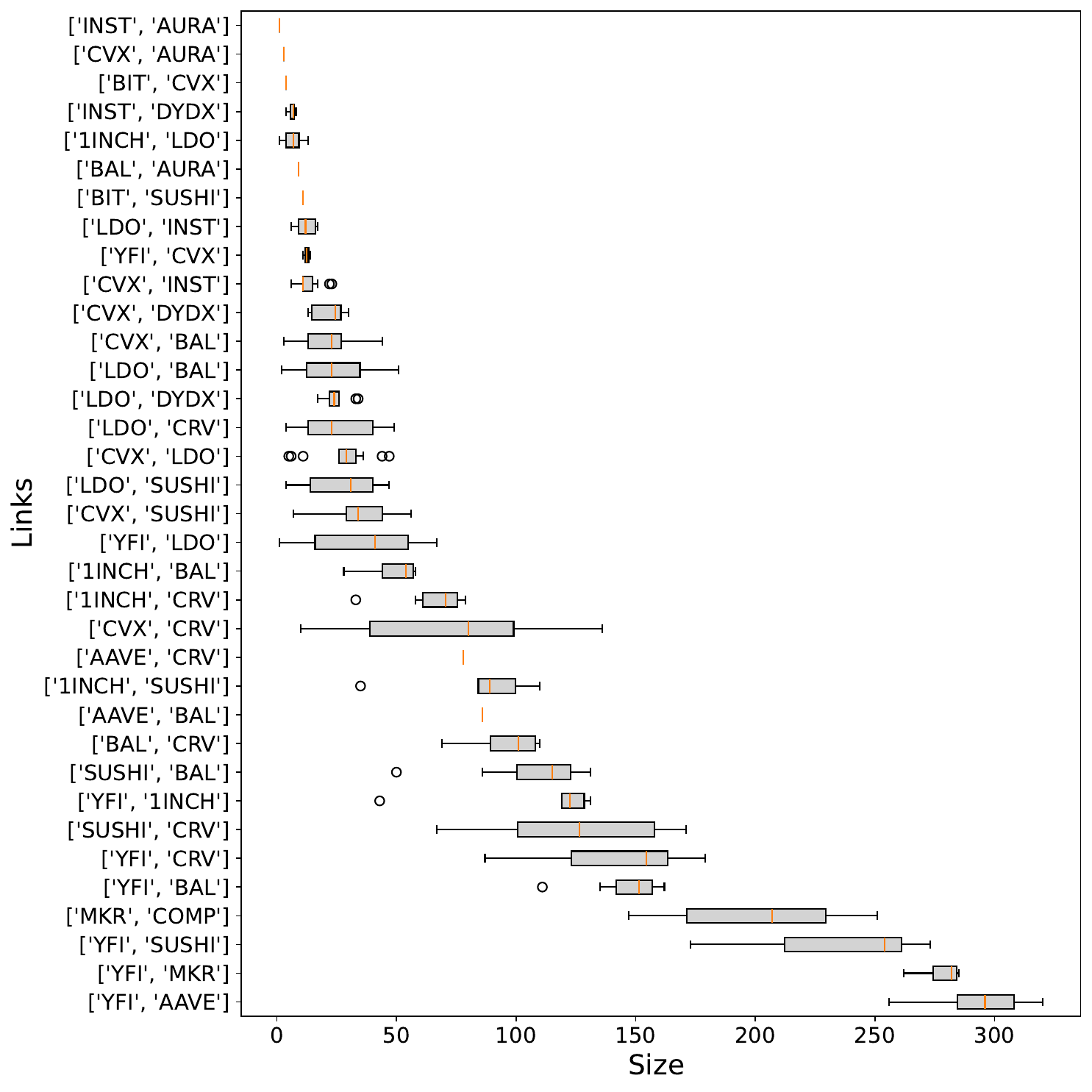}
    \caption{Distribution of the number of addresses across the sampling period that define validated links between governance tokens in validated token projections.}
    \label{fig:size_link}
\end{figure}

\paragraph{Link Size}
Figure \ref{fig:size_link} illustrates the link size counting the addresses that make up the validated links, aggregated over the sampling period. Each row represents a different pair of tokens, with the x-axis indicating the address count. 

The size of links across different token pairs shows high variability ranging from 0 to approximately 300. Some token pairs show very compact distributions indicating the consistent size of $A_{i,j}$ over the sampling period, such as 1INCH-CRV, YFI-BAL, and YFI-MKR. Other links have wide ranges, such as YFI-SUSHI, and MKR-COMP, suggesting variability in how many addresses are part of the link. It is noteworthy that the number of addresses constituting a link are small, meaning relatively few address are responsible for the formation of a validated link. 


\paragraph{Internal Influence}

The Internal Influence metric (Equation \ref{eq:internal_influence}) provides a measure of the potential governance control held by link-defining addresses within the token communities they are part of, taking into account the relative holdings of its members. By accounting for the relative holdings of these addresses, this metric highlights their capacity to exert influence if they were to act collaboratively.

\begin{figure}[H]
    \centering
    \includegraphics[width=1.05\textwidth]{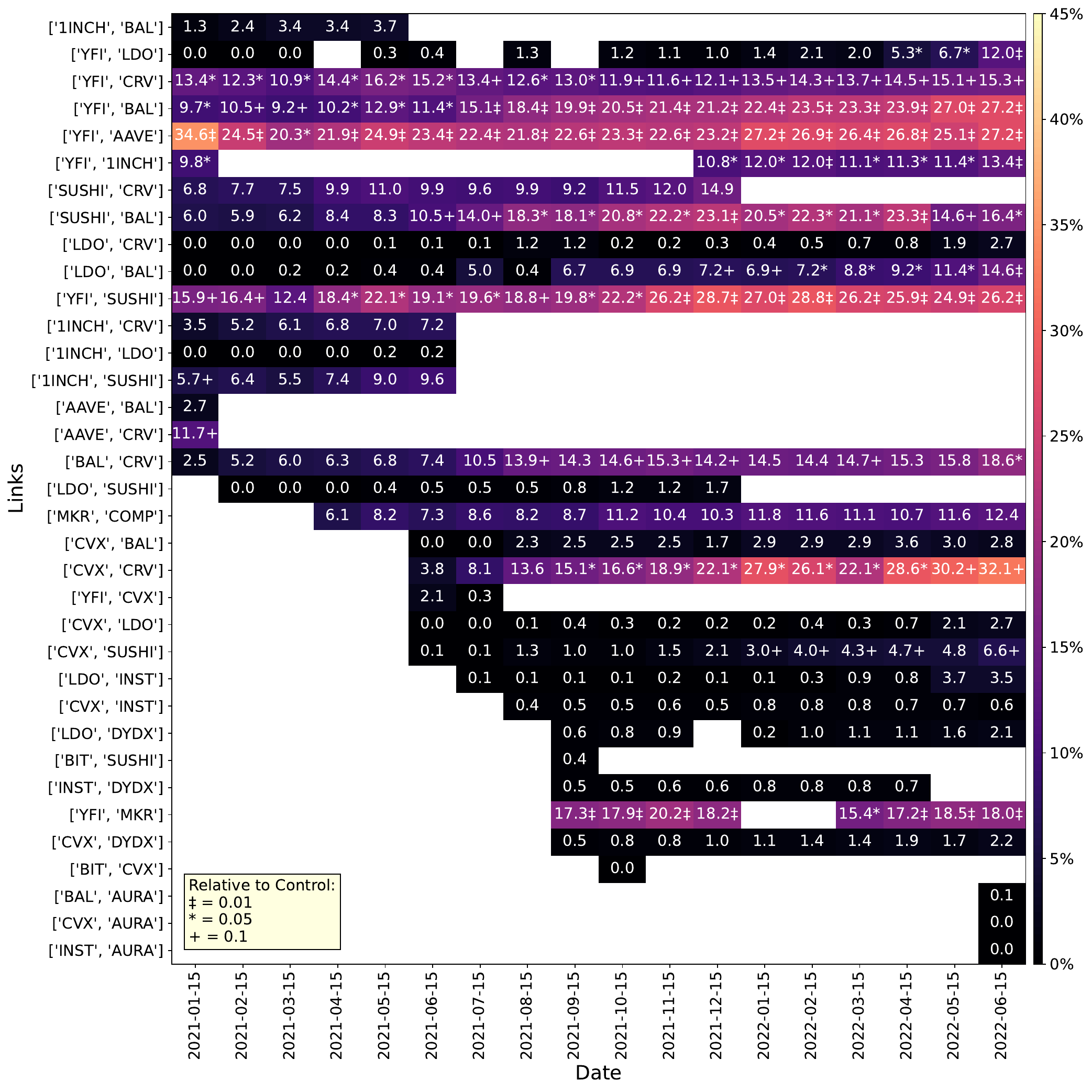}
    \caption{Influence exerted by addresses within token links.}
    \label{fig:internal_influence}
\end{figure}

In Figure \ref{fig:internal_influence} we show that across the 35 identified links, the internal influence metric ranges from close to 0 to up to 34\%. Links that consistently achieve statistical significance against the permuted control - such as YFI-CRV, YFI-BAL, YFI-AAVE, SUSHI-BAL, LDO-CRV, YFI-SUSHI, YFI-1INCH, BAL-CRV, CVX-CRV, YFI-MKR - are generally characterised by high values of internal influence. This suggests that the link-defining addresses in these pairs hold a substantial proportion of governance tokens in both communities, giving them the potential to exercise significant control.

\begin{figure}[t]
    \centering
    \includegraphics[width=0.7\linewidth]{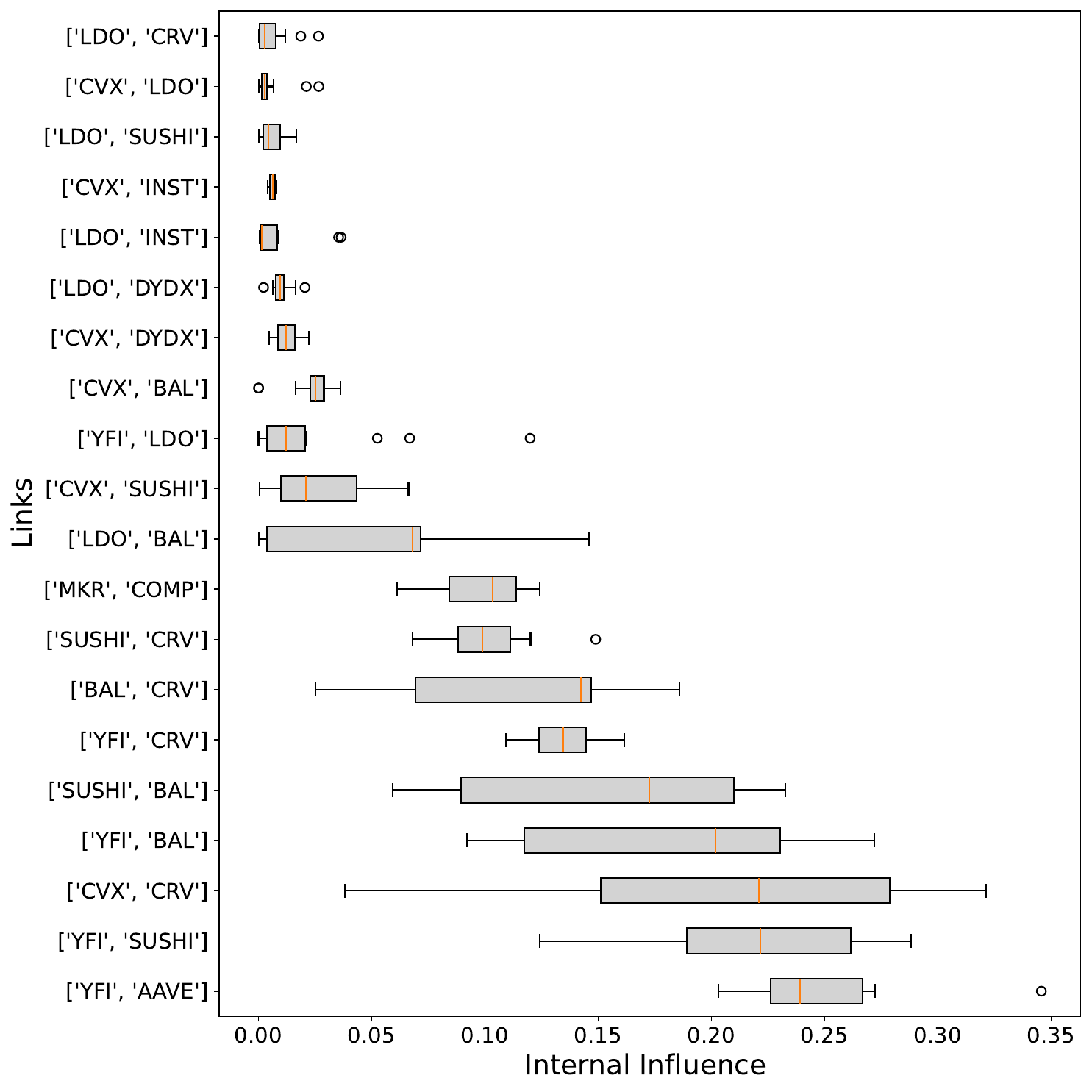}
    \caption{A collapsed time view of internal influence metrics across token links, filtered for links that occurred at least nine times.}
    \label{fig:boxplot_internal_influence}
\end{figure}

To focus on the most persistent links, we collapse the time dimension and filter for links that occurred at least nine times across the sampling period. Figure \ref{fig:boxplot_internal_influence} shows that many of these links maintain internal influence values consistently above 7\%. This indicates a high concentration of governance rights. Given the relatively small link sizes (see Figure \ref{fig:size_link}) compared to the total number of addresses holding the token (see Appendix, Figure \ref{fig:address_holding_absolute_over_time}), these rights are held by small yet influential groups.

\begin{figure}[H]
    \centering
    \includegraphics[width=0.8\linewidth]{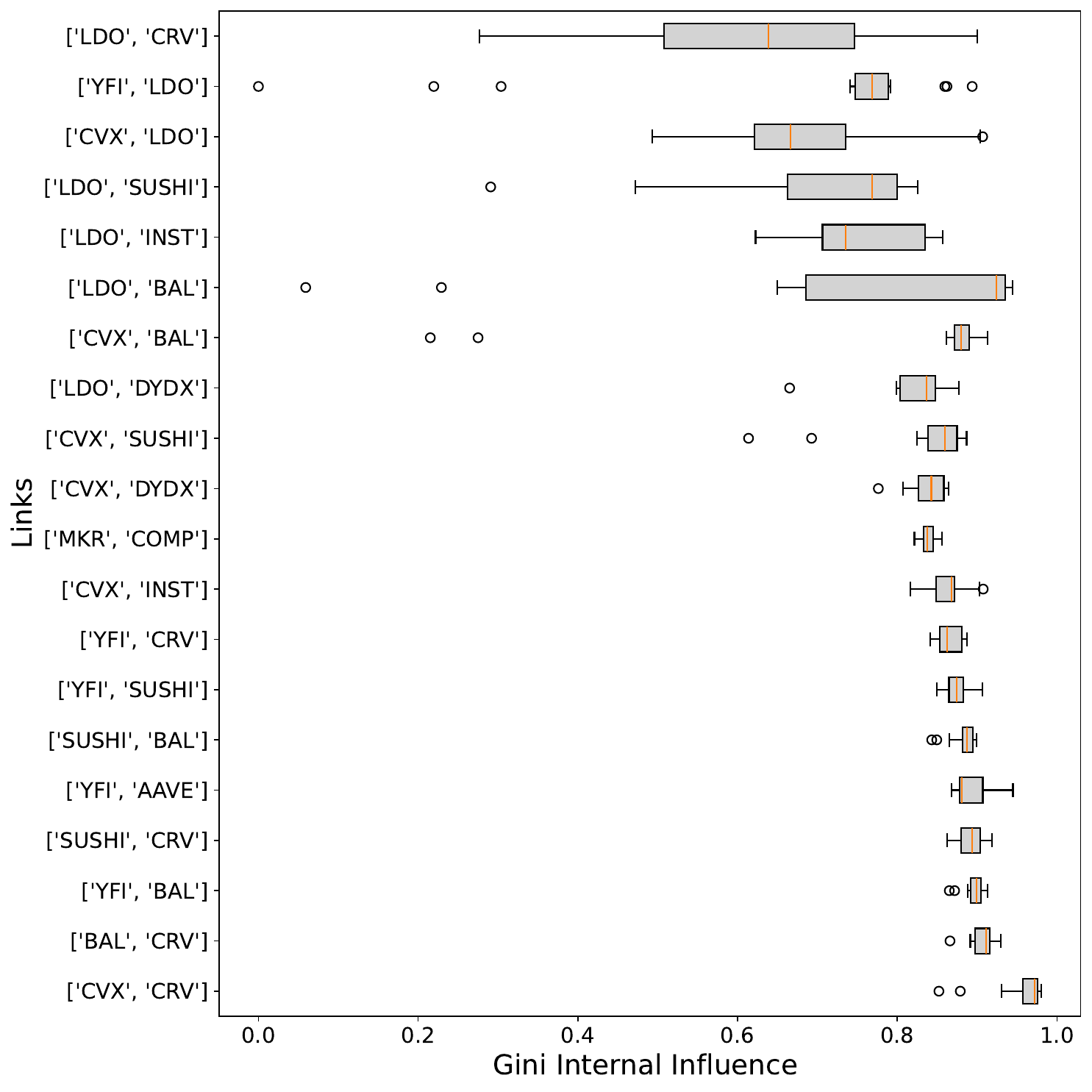}
    \caption{Gini coefficient showing the inequality of influence within token links occurring nine or more times.}
    \label{fig:gini_internal_influence}
\end{figure}

\paragraph{Gini Coefficient of Internal Influence}
To analyse the distribution of control within validated links, we examine the Gini coefficient of internal influence for link-defining addresses that formed a validated link nine or more times over the sampling period. Figure \ref{fig:gini_internal_influence} illustrates this distribution of the Gini coefficient across SVN snapshots.

We find that the influence distribution within links tends to be highly unequal with most links achieving a Gini coefficient above 0.7. This highlights that even within influential links, the distribution of influence is highly unequal, with a few meaningful addresses holding a disproportionately large share of internal influence relative to the other addresses constituting the link. This pattern seems to hold true across all of the validated links identified by the SVN filter. 

\begin{figure}[H]
    \centering
    \includegraphics[width=0.8\linewidth]{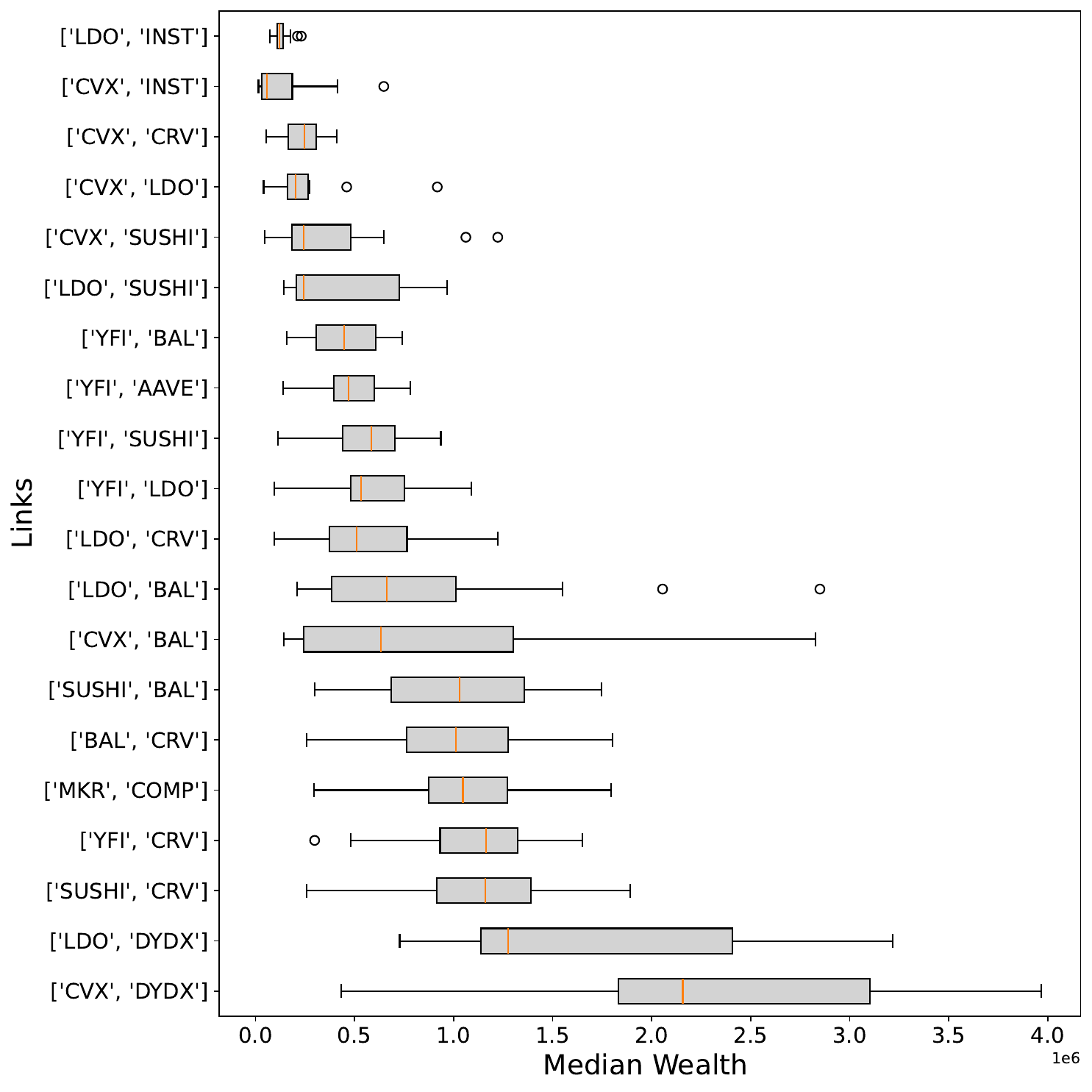}
    \caption{Median wealth held by link-defining addresses across all tokens in the sample.}
    \label{fig:median_wealth}
\end{figure}

\paragraph{Median Wealth}
Figure \ref{fig:median_wealth} shows the median wealth of link-defining addresses for links that appear at least nine times throughout the sampling period. 

We observe that the median wealth typically ranges from 200,000 USD to 1,300,000 USD. The size of the investment exceeds the typical availability of funds of households, implying that the entities controlling the addresses likely possess substantial financial resources. This may suggest that the identified entities behind the addresses are either professional investors or institutions.  

\begin{figure}[H]
    \centering
    \includegraphics[width=1\linewidth]{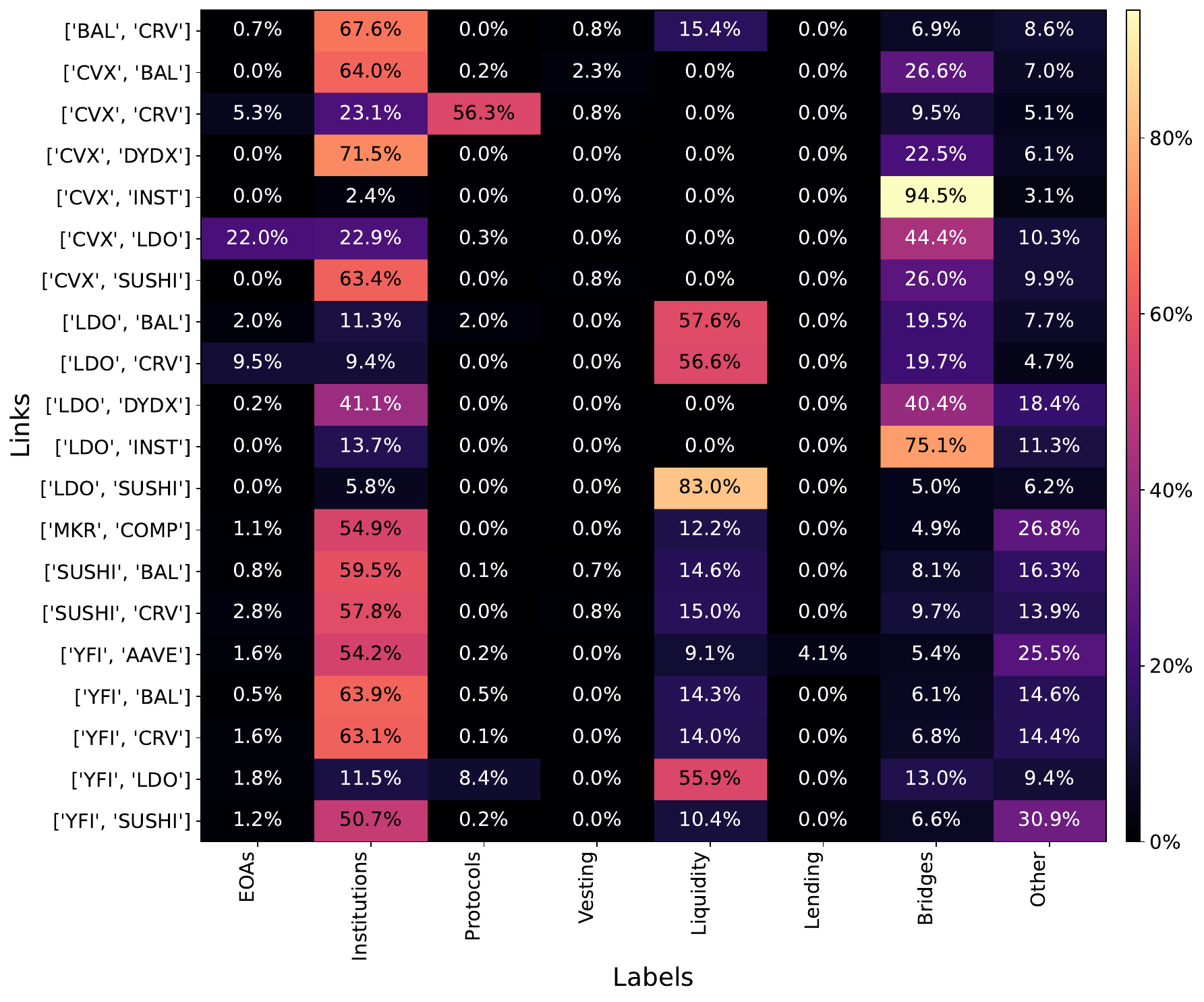}
    \caption{Relative Internal Influence of link-defining addresses by entity type.}
    \label{fig:labels_link}
\end{figure}

\paragraph{Labels of Link-defining Addresses}
We incorporate labelling to understand the composition of economic entities within validated links. Figure \ref{fig:labels_link} illustrates the relative internal influence of link-defining addresses by entity type (\ref{eq:relative_internal_influence}). We categorise labels as EOAs (individual wallets), Institutions, Protocols, Vesting Contracts, Staking Contracts, Liquidity Pool Contracts, Lending Contracts, Bridge Contracts, and Other Contracts.

The analysis reveals notable patterns across different types of links. For example, the CVX-CRV link is dominated by protocol contracts, reflecting the dependency between Convex and Curve which we explore more in the following section. Links containing the LDO token seem to form mainly from Liquidity Pool Contracts. Referring to Figure \ref{fig:size_link}, we can see that LDO Links generally have a smaller number of wallets. The INST token was minted before being distributed, which may be the reason why we see a high relative control value in bridge contracts, as addresses holding this token have been labelled as such.


However, the main result from Figure \ref{fig:labels_link} is that institutional addresses hold a majority of internal influence for most links, putting them potentially into a central role in shaping governance dynamics across token communities if they were to act on their voting rights. This aligns with the finding from Figure \ref{fig:median_wealth}, as these entities typically possess large financial resources.  

\paragraph{Directional Influence}
To understand the asymmetry of governance control within validated links, we analyse the Directional Influence Metric (Equation \ref{eq:directional_influence}). We limit the analysis to links that achieved at least one significant value in the directional control metric compared to the permuted control sample of the respective token population over the sampling period.


\begin{figure}[H]
    \centering
    \includegraphics[width=1.05\linewidth]{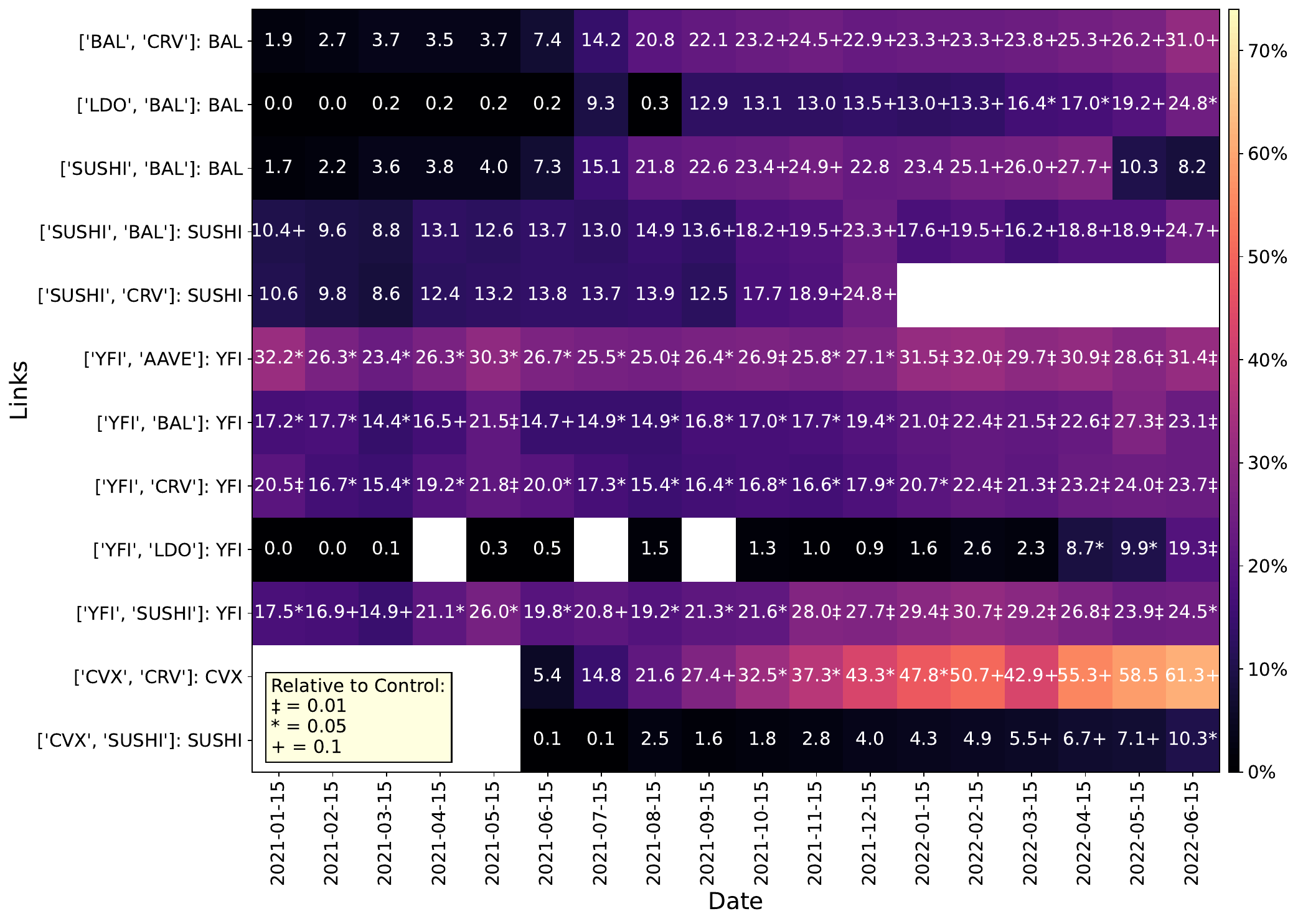}
    \caption{Directional influence within token pairs in validated links on the respective token community.}
    \label{fig:directional_influence}
\end{figure}

Figure \ref{fig:directional_influence} shows that statistically meaningful differences in directional influence values are primarily found for YFI, SUSHI, BAL and CVX. Directional influence is typically concentrated on one token within a link, with the exception of SUSHI-BAL, which suggests that link-defining addresses tend to have disproportionate control in one token community. The absence of links observed in the previous figure can be explained by control samples holding an equivalent amount or greater influence on a token.  

The directional influence on YFI by several links may be explained by its fair launch mechanism \footnote{See under \textit{Fair-Launch}: \url{https://docs.yearn.fi/resources/defi-glossary/}}. A Fair Launch coin or token is one where no pre-allocation or pre-mining is conducted for founders, developers, venture capitalists, or early investors before its public release. However, participants are still required to be aware of the launch of the token. In this context, figure \ref{fig:directional_influence} shows the connection of YFI to many established DeFi protocols that acquire directional influence over YFI, suggesting that the respective addresses already influential within existing Defi Token may were aware of the launch and hence managed to acquire a meaningful stake in the governance.

We observe that CVX-CRV link-defining addresses exert a high directional influence over CRV. It needs to be pointed out that these addresses are largely protocol contract addresses rather than EOAs (see Figure \ref{fig:labels_link}). Convex, the protocol behind the CVX governance token, optimises rewards for CRV stakers and Curve Liquidity Providers (LPs) by redirecting Curve’s token emissions and fees. Users stake CRV or lock their Curve LP tokens on Convex to earn boosted CRV rewards and additional benefits. This interdependence between Convex and Curve likely accounts for the significant directional influence of CVX on CRV \cite{convex_docs}.


Directional influence on BAL started increasing in May 2021, suggesting that link-defining addresses began accumulating the token. Archival documentation from the Balancer governance forum at the time reveals that a renewed liquidity mining program was launched in April 2021 to incentivise migration from Balancer V1 to Balancer V2, coinciding with the official launch of Balancer V2 \cite{balancer_v2_liquidity_mining}. Through this program, Liquidity Providers were able to earn BAL governance tokens as rewards for their participation in liquidity mining. This point is supported as two of the links are formed with other DEX tokens, CRV and SUSHI, suggesting the holders are familiar with DEXes. These LPs likely held onto their accumulated tokens to retain as speculative vehicles within the protocol, as a significant drop in influence is observed after April 15, 2022, reflecting the broader market downturn, with many LPs selling their BAL tokens. 

SUSHI displays more volatile directional influence patterns, possibly tied to broader market cycles, which also impact YFI and other tokens as well.

These findings highlight the asymmetric nature of governance control within validated links, with link-defining addresses often concentrating their influence within one token community. 


\subsubsection{Link Dynamics}

\paragraph{Link dynamics during Market Sell-offs}
The sampling period coincides with two major sell-offs, namely around the middle of May 2021 and the middle of April 2022 (refer to the Appendix \ref{sub-section:tvl_over_sampling}). With this information, we can explore how links and their influence change in reaction to changes in market prices to understand their holding behaviour better. 

\begin{figure}[H]
    \centering
    \includegraphics[width=0.8\linewidth]{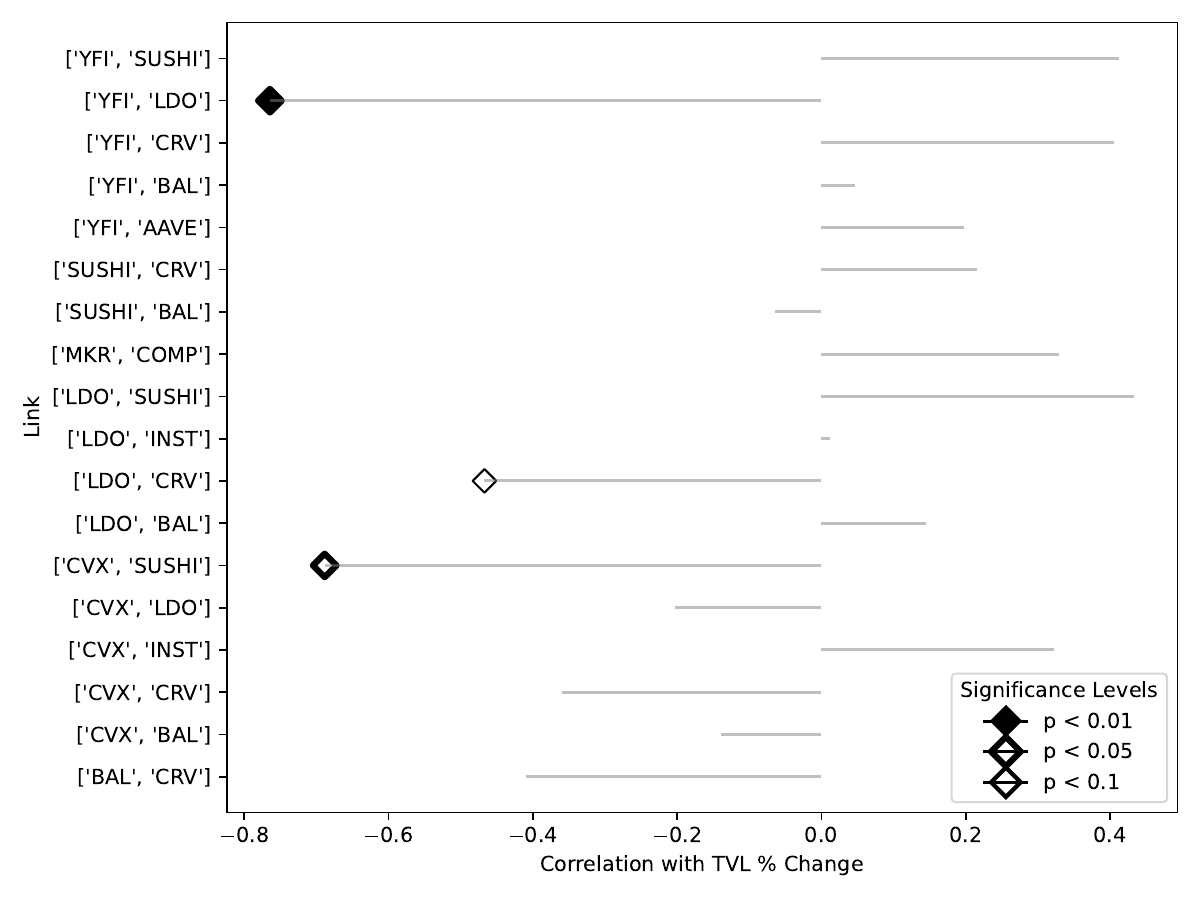}
    \caption{Correlation between Internal Influence of Link Pairs and Percentage Change in Market-wide TVL (adjusted for auto-correlation by differencing).}
    \label{fig:internal_influence_vs_tvl_correlation}
\end{figure}

In Figure \ref{fig:internal_influence_vs_tvl_correlation}, we present the correlation between the percentage change in TVL and internal influence across various link pairs over the sampling period. It illustrates the strength of these correlations, with each horizontal line representing a specific link pair. The line's length reflects the correlation's magnitude, while the dot at the end of each line indicates the statistical significance of the coefficient. 

We find that few correlations achieve statistical significance and have negative values, suggesting that as TVL increases, internal influence tends to decrease. We adjusted the values for auto-correlation by  (See Appendix Table \ref{tab:summary_stats_adjusted}). This may indicate that as speculative interest (represented by TVL) grows, the influence of core holders (represented by internal influence) becomes diluted. This relationship is significant in links like YFI-LDO, CVX-SUSHI and LDO-CRV, and highlights a potential risk of open governance models like DAOs during speculative periods, where governance might be susceptible to short-term thinking as core holders have relatively lesser influence due to the strong influx of speculators. 

\section{Discussion}
In this work, we show that DeFi governance token owners are often concentrated in a set of influential link-defining addresses, primarily classified as institutional investors, with a significant potential to exert decision-making power across multiple protocols. Validated links suggest relationships between token communities, with dominant control typically focused on one protocol. For example, the CVX-CRV overlap highlights CVX's dependency on CRV. We found correlations between internal influence and market-wide TVL, indicating that influence shifts with market cycles. This suggests that addresses constituting a link may vary as speculators dilute core holders during periods when the asset prices appreciate. 

Our findings align with prior research on DeFi governance, such as those by \cite{nadler2020decentralized} and \cite{fritsch2022analyzing}, which noted a concentration of governance control within specific token communities based on token supply. However, we move beyond these works by demonstrating how concentration extends across sets of protocols.

This has further implications. The prevalence of governance power wielded by link-defining addresses highlights potential dual incentives between entities active in both token communities. While we do not explore the actual voting behaviour of link-defining addresses in this work, we highlight the risk to the DeFi protocol of the potential existence of plurality incentives reaching beyond one protocol when making governance decisions. This is amplified by the fact that governance right a predominantly held by few institutional investor, which have a clear profit incentive. While this incentive is not inherently negative it may be at conflict with wider community interest of one or more token communities. Dominance in often one token pair making up the link may provide incentive for these actors for favouritism. This could introduce conflicting in interest when strategic choices in competition between paired tokens in a link arises. In addition, the short term dilution of key holder during price appreciation of the respective governance tokens introduces another risk, where entities treating the token as investment vehicle may take during their investment period choices which align with their investment horizon rather than with the long-term vision of the community and the project. 

Despite the fact that we were able to expand knowledge and awareness in this context, our work has some limitations. Although we attempted to quantify influence based on token holdings, it does not consistently reflect decision-making power due to variations in governance structures and voting mechanisms across protocols. In Appendix \ref{appendix:overview_governance_tokens} we provide a high-level overview of the different protocols, yet future work should consider these differences in greater detail for a more accurate assessment of the extent of influence. Also, addresses are inherently pseudonymous, and despite using triangulation methods to improve label accuracy, some mislabelling may have occurred, potentially distorting the perceived level of control of respective user groups. The integration of semi-supervised learning methods to identify addresses \citep{valadares2023mapping, beres2021blockchain} and clustering methods \citep{victor2020address} may improve our results. Lastly, our analysis focused on a limited set of governance tokens over a specific time frame. Expanding the scope to include a wider range of tokens and the duration of the observation period may capture more comprehensive trends and behaviours in links.

\section{Conclusion} 
Our research studies the underexplored risk vector of cross-protocol control in public and open token-based voting systems, such as DeFi tokens, by looking at characteristics and behaviour of addresses part of validated links in SVN responsible for introducing this control. First, we established that influential links are small in size by the number of addresses constituting them but hold meaningful influence in two protocols giving them the right to weigh and, if acted on this decision-making right, influence governance decisions in both protocols. Secondly, the addresses holding these tokens are not typical retail investors but seem to be dominated by professional and institutional investors, as indicated by the median investment size and the labelling. Thirdly, typically, a dominant token exists in a link where more influence can be exerted on one protocol compared to the respective counterpart in the link. Finally, the internal influence of link-defining addresses displays sensitivity to market cycles, with evidence of negative correlation with the TVL of its endpoint tokens. This indicates that ``core" holders, i.e. the most committed community members, may be diluted in governance power during speculative periods by trend-driven token holders. 

\section{Research Data}
All code and data are accessible on the following GitHub repository: \url{https://github.com/xm3van/reasearch-project-erc20-governance}



\clearpage
\bibliographystyle{plainnat}

\appendix
\renewcommand\thefigure{\thesection.\arabic{figure}} 
\renewcommand\thetable{\thesection.\arabic{table}}
\setcounter{figure}{0} 
\setcounter{table}{0}
\label{section:appenix}
\section{Block Heights and Snapshot Dates}\label{appendix:snapshots}
\begin{table}[H]
\centering
\begin{tabular}{|p{2cm}|p{2cm}|}
\hline
Block Height & Snapshot Date\\ 
\hline
11659570 & 2021-01-15 \\
11861210 & 2021-02-15 \\
12043054 & 2021-03-15 \\
12244515 & 2021-04-15 \\
12438842 & 2021-05-15 \\
12638919 & 2021-06-15 \\
12831436 & 2021-07-15 \\
13029639 & 2021-08-15 \\
13230157 & 2021-09-15 \\
13422506 & 2021-10-15 \\
13620205 & 2021-11-15 \\
13809597 & 2021-12-15 \\
14009885 & 2022-01-15 \\
14210564 & 2022-02-15 \\
14391029 & 2022-03-15 \\
14589816 & 2022-04-15 \\
14779829 & 2022-05-15 \\
14967365 & 2022-06-15 \\
\hline
\end{tabular}
\caption{Snapshot dates and corresponding Ethereum block heights for data collection, covering monthly intervals from January 15, 2021, to June 15, 2022.}\label{tab:sample_intervals}
\end{table}

\section{Token Ownership Over Time}
\setcounter{figure}{0} 
\setcounter{table}{0}

\begin{figure}[H]
    \centering
    \includegraphics[width=\linewidth]{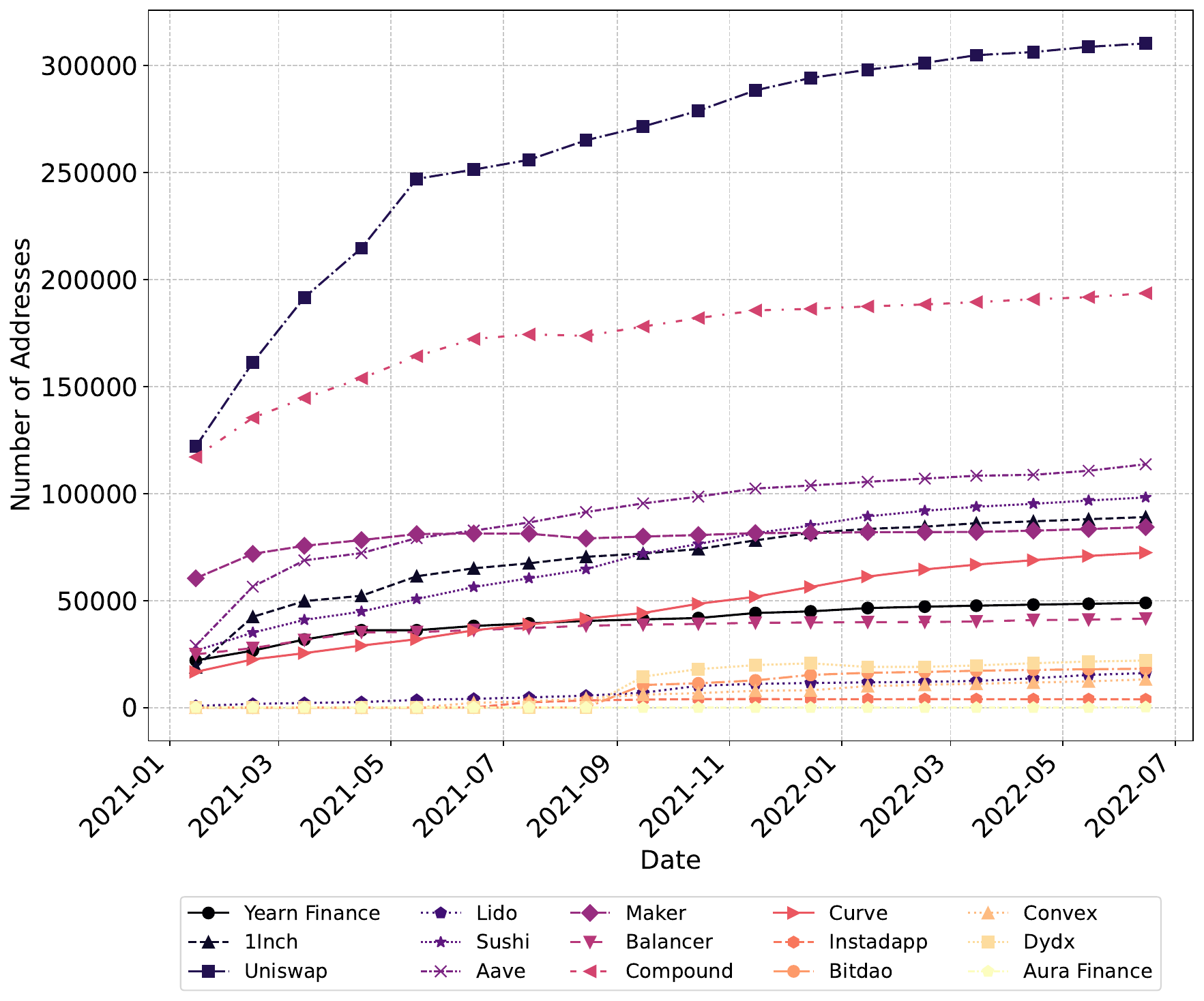}
    \caption{Absolute number of addresses holding each governance token over time.}    \label{fig:address_holding_absolute_over_time}
\end{figure}

Figure \ref{fig:address_holding_absolute_over_time} shows the absolute number of addresses holding governance tokens over time. The introduction of new tokens after June 2021 contributes to the structural break observed in the Jaccard Similarity Matrix (Figure \ref{fig:jaccard_similarity}).

\newpage
\section{TVL changes over the Sampling Period}\label{sub-section:tvl_over_sampling}
\setcounter{figure}{0} 
\setcounter{table}{0}

\begin{figure}[H]
    \centering
    \includegraphics[width=0.9\linewidth]{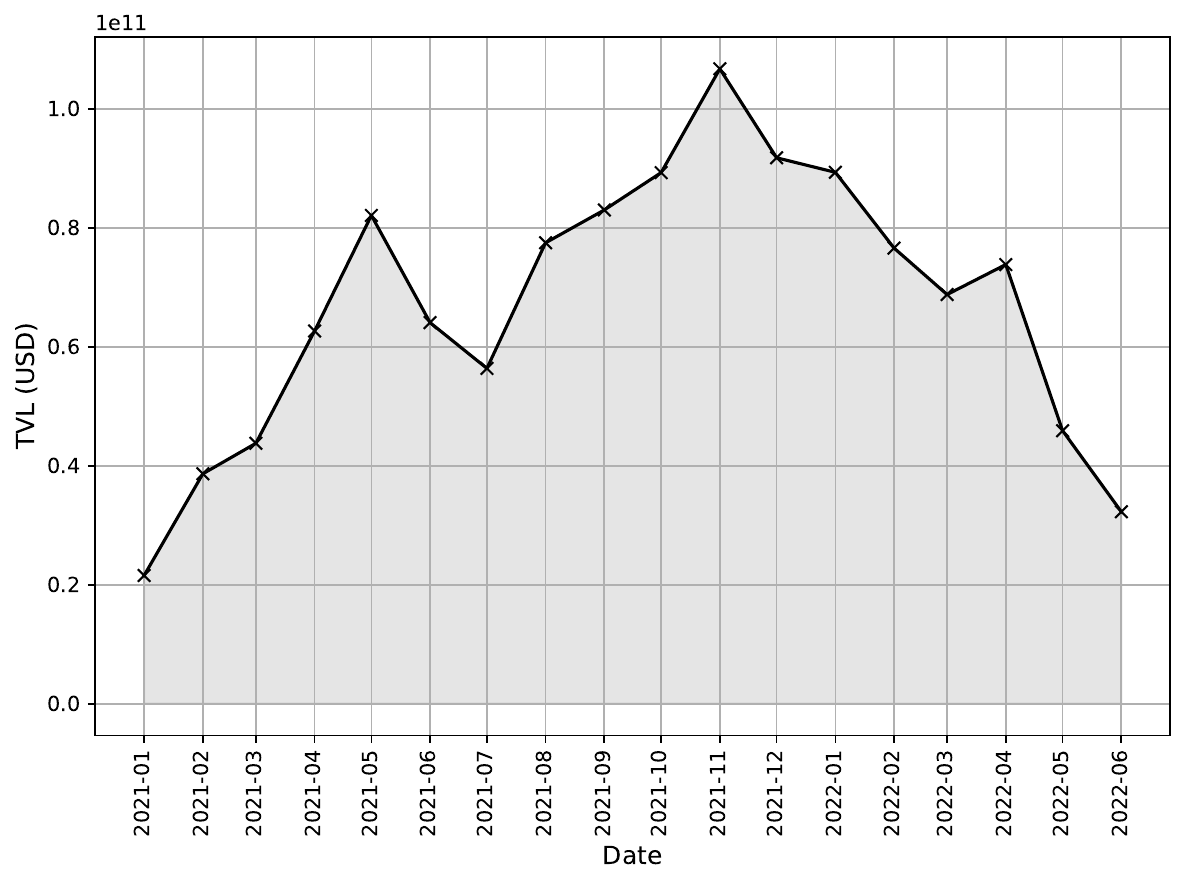}
    \caption{The figure shows the dollar-denominated total value locked (TVL) in Ethereum-based DeFi protocols over time. The chart illustrates market cycles, with significant declines in May 2021 and April 2022, corresponding to broader market sell-offs.}
    \label{fig:tvl}
\end{figure}

Figure \ref{fig:tvl} shows the dollar-denominated Total Value Locked (TVL) in Ethereum DeFi-Protocols. Due to the dollar denomination, it follows market cycles with the depreciation of cryptoassets in a downturn.

\section{Summary Statistics for Correlation and Autocorrelation Analysis}\label{sub-section:autocorr}
\setcounter{figure}{0} 
\setcounter{table}{0}

\begin{table}[H]
\begin{tabular}{lrrrr}
\toprule
 & Correlation & P-value & Durbin-Watson & Ljung-Box P-value \\
Link Name &  &  &  &  \\
\midrule
\textit{['BAL', 'CRV']} & -0.4095 & 0.1026 & 0.7388 & 0.8041 \\
\textit{['CVX', 'BAL']} & -0.1387 & 0.6674 & 2.1857 & 0.2479 \\
\textit{['CVX', 'CRV']} & -0.3596 & 0.2510 & 1.2972 & 0.5627 \\
\textit{['CVX', 'INST']} & 0.3227 & 0.3631 & 2.3978 & 0.5616 \\
\textit{['CVX', 'LDO']} & -0.2033 & 0.5263 & 0.8803 & 0.6700 \\
\textit{['CVX', 'SUSHI']} & -0.6882 & 0.0134 & 1.0180 & 0.5627 \\
\textit{['LDO', 'BAL'] }& 0.1443 & 0.5806 & 2.6991 & 0.2802 \\
\textit{['LDO', 'CRV'] }& -0.4669 & 0.0588 & 1.3538 & 0.5478 \\
\textit{['LDO', 'INST'] }& 0.0113 & 0.9737 & 2.0899 & 0.9814 \\
\textit{['LDO', 'SUSHI'] }& 0.4333 & 0.2110 & 1.0335 & 0.6265 \\
\textit{['MKR', 'COMP'] }& 0.3286 & 0.2514 & 2.1958 & 0.4494 \\
\textit{['SUSHI', 'BAL']} & -0.0649 & 0.8045 & 2.2561 & 0.2765 \\
\textit{['SUSHI', 'CRV']} & 0.2158 & 0.5239 & 1.4347 & 0.7205 \\
\textit{['YFI', 'AAVE']} & 0.1974 & 0.4475 & 0.9116 & 0.8980 \\
\textit{['YFI', 'BAL']} & 0.0469 & 0.8581 & 1.5453 & 0.3596 \\
\textit{['YFI', 'CRV']} & 0.4047 & 0.1071 & 1.5802 & 0.1811 \\
\textit{['YFI', 'LDO'] }& -0.7641 & 0.0062 & 0.7247 & 0.3374 \\
\textit{['YFI', 'SUSHI']} & 0.4116 & 0.1007 & 2.1141 & 0.8598 \\
\bottomrule
\end{tabular}
\caption{Summary Statistics for Correlation and Autocorrelation Analysis}
\label{tab:summary_stats_adjusted}
\end{table}

\newpage
\section{Sensitivity Analysis}
\setcounter{figure}{0} 
\setcounter{table}{0}

\begin{figure}[H]
    \centering
    \includegraphics[width=\textwidth]{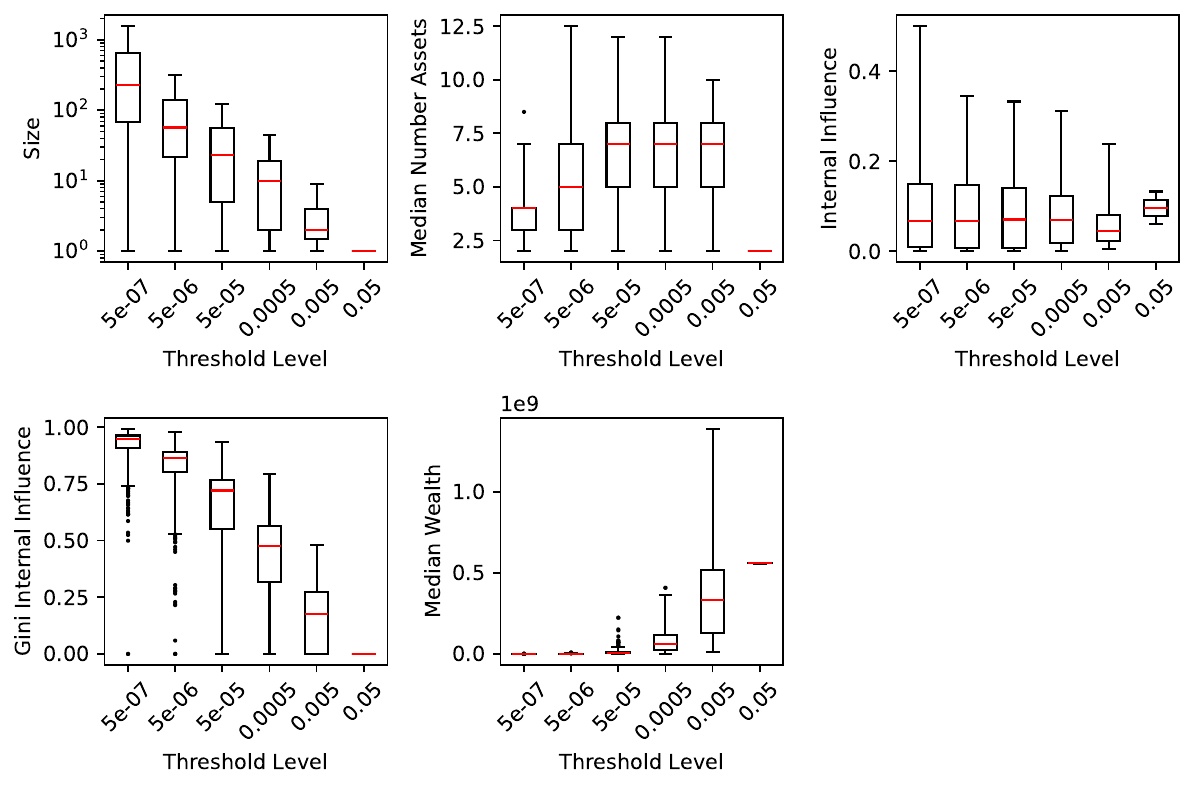}
    \caption{Sensitivity analysis of the inclusion threshold for link-defining wallets based on relative percentage supply held. The analysis evaluates the stability of the results across varying thresholds, highlighting that metrics for the chosen threshold of 0.0005\% of the token supply are relatively insensitive to higher lower thresholds.}
    \label{fig:sensitivity_analysis}
\end{figure}

We conducted the sensitivity analysis on the supply threshold for the inclusion of link-defining wallets. In this analysis, we computed the minimum, maximum, and interquartile range. The threshold level throughout the analysis is 5e-06, meaning we consider all addresses that hold tokens with at least 0.0005\% of the available supply of any token in our sample. The sensitivity analysis, depicted in Figure \ref{fig:sensitivity_analysis}, shows that results tend to be relatively stable for the metrics utilised in this research. 

\newpage
\section{Overview Governance Token}\label{appendix:overview_governance_tokens}
\setcounter{figure}{0} 
\setcounter{table}{0}

Governance tokens and the rights associated with them are unique to each protocol, giving varying degrees of control and responsibility to token holders. To provide the reader with an idea of what governance rights entail we reviewed the protocol documentation.  

\textbf{Uniswap:} Uniswap is a decentralised exchange protocol. It is governed by UNI token holders. The holder generally delegates to professional participants (either companies, university clubs or individuals) that are supposed to represent their vote.  Decisions revolve around a broad range of topics such as fee parameters \citep{uniswap_whitepaper}, delegation to different stakeholders, new protocol initiatives and features \citep{uniswap_forum}. The Uniswap governance process begins with proposal creation, requiring 2.5 million UNI tokens. First, a "Temperature Check" gauges community interest via a forum poll with at least 25,000 UNI tokens needed to start. If successful, a "Consensus Check" follows, requiring 50,000 UNI in a formal Snapshot vote. The final stage is the on-chain "Governance Proposal," which undergoes a 7-day vote, needing 40 million UNI to pass and be executed. This structured process ensures thorough community involvement in decision-making \citep{uniswap_docs}.

\textbf{MakerDAO:} MakerDAO is a decentralised credit platform that allows users to generate DAI, a stablecoin pegged to the US dollar, by locking up collateral. MakerDao Governance scope can be summarised with five areas: Stability, which centres on the financial stability and the Dai stablecoin; Accessibility, which targets frontends and distribution; Protocol, dedicated to technical development, maintenance, and security; Support, aimed at ecosystem support through tools and activities; and Governance, which deals with the interpretation of Alignment Artifacts and the balance of powers \citep{makerdao_mip101}. Governance occurs through both off-chain and on-chain processes, with proposals discussed in the Maker Governance Forum before being voted on-chain. MKR holders have proportional voting power based on the amount of MKR they hold. It can be delegated to different users. The MKR token also acts as a backstop mechanism if credit becomes under-collateralised pushing MKR holders for poor governance. \citep{makerdao_voting}.

\textbf{Aave:} Aave is a decentralised lending protocol where users can borrow and lend assets. Its governance is managed by AAVE token holders. The scope of governance can be summarised as enabling upgrades to governance itself, operational processes, voting on asset listings and risk parameters and treasury allocation (e.g. hiring, incentives) \citep{aave_governance} Proposals are discussed in public forums. Once a proposal garners enough support it is formalised into Aave Improvement Proposals (AIPs) and voted on. Successful proposals are implemented to enhance the protocol's functionalities and security. The protocol also includes a staking mechanism where AAVE holders earn rewards and help secure the protocol. The Aave governance structure allows risk admins to adjust risk parameters without requiring a vote for every change \citep{aave_governance_process}. 

\textbf{Lido:} Lido is a liquid staking protocol that enables users to stake assets and remain liquid by issuing a derivative token. Governed by LDO token holders, their responsibilities include approving smart contract upgrades, managing the treasury, setting staking parameters, and overseeing node operator management—this covers the onboarding, evaluation, and potential replacement of operators. They also allocate grants for community and development projects through the Lido Ecosystem Grants Organisation. The governance process engages the community in initial discussions and feedback on the research forum, followed by a Snapshot vote requiring participation from over 5\% of LDO holders. If necessary, proposals move to an on-chain vote using the Aragon framework, needing a quorum of 5\% and 50\% approval to pass. In urgent situations, LDO holders can bypass standard procedures to make swift decisions. \citep{lido_docs}

\textbf{Yearn Finance:} Yearn Finance is a yield optimisation protocol that aggregates yields from various DeFi platforms. It operates under a multi-DAO structure managed by constrained delegation. YFI holders are primarily responsible for proposing, discussing, and implementing changes through three main types of proposals: Yearn Improvement Proposals (YIPs), are formal proposals that can execute any power delegated to YFI holders or address issues outside the predetermined scope of powers; Yearn Delegation Proposals (YDPs), this type allows for the redistribution of decision-making powers among different operational teams within the ecosystem; Yearn Signalling Proposals (YSPs), are used to gauge community sentiment on various topics and are non-binding. YFI holders can: Manage and reallocate discrete powers among yTeams, specialised groups focusing on different aspects of the protocol, interact with the YFI token contract, including actions like minting new tokens or burning existing ones, setting and adjusting fee structures across the Yearn Protocol, select or change signatories of the multisig wallet, which holds significant operational powers including the execution or vetoing of on-chain decisions, ratify or deratify yTeams, thereby influencing which teams hold delegated powers, allocate and manage funds from Yearn’s treasury. \citep{yearn_operations}

The governance flow typically involves yTeams proposing decisions to a transactional team (yTx), which then creates delegated transactions sent to the Multisig for execution or veto. This structured yet dynamic governance framework allows YFI holders to effectively oversee and direct the continuous development of the Yearn protocol. \citep{yearn_proposal_process}

\textbf{Aura Finance:} Aura Finance is a protocol that aims to increase yields on Balancer liquidity pools. AURA token holders can lock their tokens to obtain veAURA, enhancing both their voting power and rewards, with incentives structured to favour long-term engagement up to 16 weeks. The voting power is proportionate to the amount of veAURA held \citep{aura_vote_locking}. Governance participation is enabled by holding veAURA, which grants the right to propose and vote fee adjustments, tokenomics, and strategic initiatives \citep{aura_governance}. 

\textbf{BitDAO:} BitDAO is a decentralised autonomous organisation that supports DeFi projects through grants and partnerships, it now pivoted to become Mantle a product suite offering different decentralised technologies. BIT token holders engage in governance via delegated voting, allowing them to vote on proposals or delegate their voting power to others. Governance activities are primarily conducted through the Snapshot platform, ensuring transparency and efficiency by aggregating votes off-chain with potential future shifts to on-chain mechanisms. The proposal can cover a wide range of decisions from operational changes, and treasury management, to strategic initiatives, and requires sufficient community backing to be formalised into official votes. A multi-sig wallet executes these decisions. \citep{bitdao_docs}.

\textbf{SushiSwap:} SushiSwap is primarily a decentralised exchange protocol. It offers additional financial tools such as yield-generating instruments, bonds, and a platform for token streaming and vesting. SushiSwap's governance is powered by its community through a combination of forum discussions and Snapshot voting. Proposals can be submitted by any community member and, if they gather sufficient interest, are formalised through Snapshot where they are voted upon. The SUSHI token gives the holder voting power only if it is staked within a protocol-specific staking contract or deposited in the SUSHI-ETH pool. The process is underpinned by a multi-signature mechanism that involves prominent members from the DeFi community who execute or veto decisions based on the collective voting outcomes. \citep{sushiswap_docs}

\textbf{dYdX:} The dYdX is a decentralised derivatives exchange, since the study period has concluded the protocol has moved from Ethereum. DYDX token holders were responsible for managing proposals affecting both strategic and operational aspects of the protocol, including key protocol amendments, liquidity, safety modules, and reward distributions. Proposals, categorised by their impact, pass through a lifecycle involving community discussion, off-chain drafting, on-chain voting, and execution via time-locked contracts. DYDX holders can delegate voting power on their behalf \citep{dydx_docs}.

\textbf{Instadapp:} Instadapp is a middleware that aggregates multiple DeFi protocols into one upgradable smart contract layer. It uses the INST token for governance using a similar structure to Compound thus allowing delegation. Token holders can propose changes to the protocol, vote on upgrades, and influence the management of community treasury funds. \citep{instadapp_protocol_governance} Governance starts with discussions in community forums, followed by an off-chain vote via snapshot followed by an on-chain vote via atlas. \citep{instadapp_governance_flow}.

\textbf{Curve:} Curve is a decentralised exchange protocol, stablecoin provider, and lending platform on Ethereum and EVM-compatible chains. Curve's governance is managed by CRV token holders who can lock their tokens to receive veCRV, which grants voting power. Users need at least 2,500 veCRV to create proposals, while anyone can vote with no minimum required. Proposals have a voting duration of seven days, and voting power decays linearly over time. Curve utilises three types of votes: ownership votes, parameter votes, and emergency votes, each with specific quorum requirements. The EmergencyDAO, a group of trusted agents, can intervene in critical situations, such as shutting down liquidity pools or gauges. The scope of governance ranges from incentive allocation on gauges to allocating future CRV emissions to liquidity pools, protocol upgrades and allocation of treasury and management. \citep{curve_docs}

\textbf{Convex:} Convex Finance is built on top of different DeFi protocols utilising a locked voting mechanism which enables the allocation of future token emissions (e.g. Curve, Frax Finance). It allows liquidity providers and stakers to earn boosted rewards without needing to lock their tokens directly in the underlying protocols. By aggregating user stakes, Convex optimizes yield for users while offering its own native token (CVX) as an additional incentive. CVX, in addition, is used to vote on proposals. To participate in governance voting, users must lock their CVX tokens for a minimum of 16 weeks, creating vote-locked CVX with governance power, similar to Curve. This lock-in period aligns with the protocol's weekly epoch cycles. Locked CVX also accumulate rewards based on the protocol's earnings, providing financial incentives alongside governance influence. Convex's governance is further reinforced by a multi-sig that ensures that all CVX votes align with the interests of the protocols involved, blocking any harmful proposals. \citep{convex_docs}

\textbf{Balancer:} Balancer is a decentralised exchange protocol that allows for customizable liquidity pools and decentralised trading. Governance is handled through veBAL, where token holders lock BAL/WETH Balancer Pool Tokens for up to 1 year to participate in governance.  Token holders can influence key decisions regarding protocol fees, and liquidity incentives by directing emissions of BAL, and other allocations of treasury resources to projects. The governance process begins with a decision on the forum, being formalised in a proposal, voted on and then implemented on-chain by a multisig arrangement of trusted members. The governance process is supplemented by an Emergency subDAO that can shut down or pause certain function-critical contracts. \citep{balancer_docs}.

\textbf{Compound:} Compound allows users to borrow and lend cryptocurrencies through a decentralised market. Users can earn interest on their deposits or take out loans against their crypto assets. The governance of Compound is managed by COMP token holders. COMP holders can delegate their vote or vote themselves. The governance process generally starts with a discussion on the forum, any address with more than 25 000 COMP delegates can vote on put a proposal forward. A proposal must achieve a majority of the votes with a minimum quorum of 400,000 COMP votes for approval. Approved proposals are queued in a Timelock contract for two days before implementation. The scope of governance includes adjustments to interest rates, adding or removing supported assets, protocol upgrades, and changes to the risk parameters. COMP holders have significant influence over the protocol’s evolution and operational decisions, ensuring a decentralised and user-driven governance process. \citep{compound_docs}.

\textbf{1Inch:} 1inch operates as a decentralised exchange aggregator that optimises trades across multiple DEXes. The governance of 1inch is managed through a DAO, where 1INCH token holders can participate in governance decisions. To vote users need to either stake in the protocol’s governance contract. User can increase their power by locking their token for a longer time. Votes can be delegated, both on-chain and off-chain. The scope of governance for token holders includes proposing and voting on protocol upgrades, fee changes, and feature additions \citep{1inch_docs}. 

\end{document}